\documentclass{article}
\usepackage{spconf,amsmath,graphicx,hyperref}

\usepackage{amsfonts}
\usepackage{amssymb}
\usepackage{subcaption}
\usepackage{siunitx}
\usepackage{booktabs}
\usepackage{tikz}
\usepackage{xcolor}
\usepackage{adjustbox}
\usetikzlibrary{arrows.meta, positioning, calc, shapes}
\usepackage{multirow}

\title{Training Flow Matching Models with Reliable Labels via Self-Purification}
\name{Hyeongju Kim \qquad Yechan Yu \qquad June Young Yi \qquad Juheon Lee\thanks{Corresponding author: \texttt{hyeongju@supertone.ai}}}
\address{Supertone, Inc.}

\begin{document}
\ninept
\maketitle
\begin{abstract}
Training datasets are inherently imperfect, often containing mislabeled samples due to human annotation errors, limitations of tagging models, and other sources of noise. Such label contamination can significantly degrade the performance of a trained model. In this work, we introduce Self-Purifying Flow Matching (SPFM), a principled approach to filtering unreliable data within the flow-matching framework. SPFM identifies suspicious data using the model itself during the training process, bypassing the need for pretrained models or additional modules. Our experiments demonstrate that models trained with SPFM generate samples that accurately adhere to the specified conditioning, even when trained on noisy labels. Furthermore, we validate the robustness of SPFM on the TITW dataset, which consists of in-the-wild speech data, achieving performance that surpasses existing baselines.

\end{abstract}
\section{Introduction}
\label{sec:introduction}

The quality of training data is crucial for the performance of deep learning models. In particular, conditional generative models are highly sensitive to the fidelity and accuracy of the labels that guide their generative process~\cite{na2024labelnoise, chen2023labelretrievalaugmented, hou2025directional, cong2025guiding}. When trained on precisely annotated data, these models can learn the desired conditional distribution and generate samples that faithfully adhere to specified conditions. On the other hand, noisy or mislabeled data can severely degrade performance, leading to poor generalization and unreliable conditional generation~\cite{thekumparampil2018robustness, kaneko2019label}. Therefore, minimizing label noise is essential when training conditional generative models.

Despite its critical importance, ensuring high-quality labels remains a challenging problem. Human annotation is inherently error-prone, frequently producing mislabeled samples due to fatigue, ambiguity, or subjective interpretation~\cite{commonvoice:2020, sylolypavan2023impact}. Similarly, automated labeling methods, such as model-based tagging, can introduce biases and inaccuracies, particularly when applied to out-of-domain data~\cite{jung25c_interspeech, radford2023robust}. The growing scale of modern deep learning datasets further compounds this issue, as maintaining consistent annotation quality across such large datasets becomes increasingly difficult~\cite{he2024emilia}. These factors collectively contribute to label contamination, which poses a significant obstacle to training models on reliable data.

To address this issue, we introduce \textbf{Self-Purifying Flow Matching (SPFM)}, a novel method designed to automatically filter unreliable samples during the training of conditional generative models. Specifically, SPFM leverages the difference between conditional and unconditional flow-matching losses as a criterion for identifying untrustworthy data~\cite{lipman2023flow, dhariwal2021diffusion, ho2021classifier}. This formulation enables the model to recognize suspicious labels during training, without relying on any pretrained models, preprocessing pipelines, auxiliary objectives, or additional modules. Moreover, SPFM can be seamlessly integrated with conditional flow-matching models, providing a principled and effective mechanism for enhancing dataset quality on-the-fly.

We validate the effectiveness of our method through both controlled and real-world experiments. In synthetic two-dimensional datasets with noisy labels, we demonstrate that flow-matching models trained with SPFM generate samples that faithfully adhere to the specified conditioning, whereas standard flow-matching models struggle to recover the noise-free target distribution. Beyond these synthetic settings, we evaluate our approach on the TITW dataset~\cite{jung25c_interspeech}, which contains in-the-wild speech data with substantial noise and presents a significant challenge for many conventional text-to-speech (TTS) models. Our experiments show that the application of SPFM to SupertonicTTS~\cite{kim2025supertonictts} not only enhances its robustness but also establishes a new benchmark, outperforming existing TTS models.

\section{Related Work}
Training conditional generative models under label noise has been extensively studied. For example, transition-aware weighted denoising score matching (TDSM)~\cite{na2024labelnoise} introduces a weighted objective to mitigate the effects of noisy labels. This approach, however, requires an additional time-dependent noisy-label classifier to compute the transition-aware weights, increasing training complexity. The label-retrieval-augmented (LRA)~\cite{chen2023labelretrievalaugmented} diffusion model constructs pseudo-clean labels through neighbor consistency. However, it relies on two pretrained encoders, one for estimating category means and another for extracting high-level features, which increases computational cost and dependency on external modules. Directional label diffusion (DLD)~\cite{hou2025directional} decouples the diffusion process into random and directional paths using a dual diffusion architecture, also requiring a pretrained feature extractor during training. Finally, score-based discriminator correction (SBDC)~\cite{cong2025guiding} provides inference-time guidance for pretrained conditional diffusion models by employing an additional discriminator to distinguish correct from incorrect labels. However, it does not modify the training procedure and cannot prevent instability or performance degradation when models are trained on a noisy dataset. 

In contrast, SPFM introduces a simple and efficient training-time data selection strategy that leverages the difference between conditional and unconditional losses for each sample. SPFM operates entirely within the training phase and requires no external modules or pretrained networks. While most existing approaches attempt to reassign or correct noisy labels, SPFM selectively filters out unreliable labels and uses the corresponding samples solely for unconditional training. This approach provides a more streamlined and flexible solution for training under label noise.

\section{Flow Matching}
Flow matching~\cite{lipman2023flow} is a generative modeling framework that estimates a vector field to transport a simple base distribution to a complex target data distribution. Formally, given a base distribution $p_0(\mathbf{x})$ (e.g., standard Gaussian) and a target distribution $p_1(\mathbf{x})$, flow matching aims to learn a time-dependent vector field $v(\mathbf{x},t)$ that transforms samples $\mathbf{x}_0 \sim p_0(\mathbf{x})$ into samples $\mathbf{x}_1 \sim p_1(\mathbf{x})$ as time evolves from $t=0$ to $t=1$.

\subsection{Flow matching loss}
In practice, flow-matching models are trained to minimize the discrepancy between the model-predicted velocity and the ground-truth velocity defined on a per-sample basis along the interpolation between $\mathbf{x}_0$ and $\mathbf{x}_1$. In the conditional generative setting, where each sample is associated with some context $\mathbf{c}$ (e.g., class labels or text), the flow-matching loss $\mathcal{L}_{\text{FM}}(\theta)$ is given by
\begin{equation}
\label{eq:flow}
\mathcal{L}_{\text{FM}}(\theta) = \big\| \mathbf{v}_\theta(\mathbf{x}_t, t, \mathbf{c}) - (\mathbf{x}_1 - \mathbf{x}_0) \big\|^2,
\end{equation}
where $\mathbf{v}_\theta$ denotes the model-predicted velocity, and $\mathbf{x}_t = (1-t)\mathbf{x}_0 + t\mathbf{x}_1$ is the interpolated point along the trajectory from $\mathbf{x}_0$ to $\mathbf{x}_1$.

\subsection{Classifier-free guidance} 
To improve conditional generation quality, flow-matching models often employ \textit{classifier-free guidance} (CFG)~\cite{ho2021classifier}, which employs an unconditional mode by randomly dropping the condition $\mathbf{c}$ during training. In this setup, the model learns both conditional and unconditional vector fields, represented as $\mathbf{v}_\theta(\mathbf{x}_t, t, \mathbf{c})$ and $\mathbf{v}_\theta(\mathbf{x}_t, t, \varnothing)$, respectively, where $\varnothing$ denotes the absence of conditioning. The unconditional vector field is trained with the same objective as Eq.~\ref{eq:flow}, but without the condition $\mathbf{c}$. During sampling, the unconditional velocity guides the conditional generation as follows:
\begin{equation}
\mathbf{v}_{\text{guided}} = (1+\omega_\text{cfg})\mathbf{v}_\theta(\mathbf{x}_t, t, \mathbf{c}) 
- \omega_{\text{cfg}} \mathbf{v}_\theta(\mathbf{x}_t, t, \varnothing),
\end{equation}
where $\omega_{\text{cfg}} > 0$ is a guidance scale that controls the strength of conditioning. This approach encourages generation that more closely aligns with the desired context $\mathbf{c}$.

\section{Proposed method}

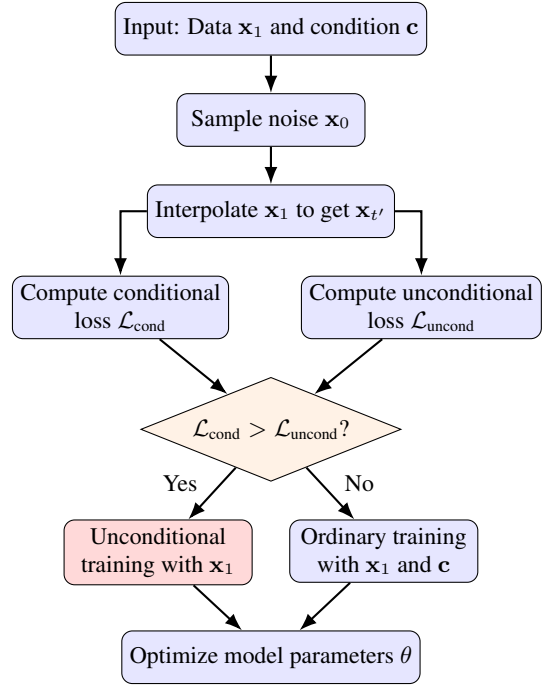
\begin{figure}[t]
    \centering
    \begin{adjustbox}{max size={\columnwidth}{\textheight}}
        \begin{tikzpicture}[
            node distance=0.5cm,
            block/.style={rectangle, draw, rounded corners, minimum height=0.7cm, minimum width=2.5cm, align=center, fill=blue!10, font=\small},
            decision/.style={diamond, draw, aspect=2.5, minimum height=1.0cm, align=center, fill=orange!10, font=\small},
            line/.style={-Latex, thick},
            label style/.style={font=\tiny}
        ]

        % Nodes
        \node (start) [block] {Input: Data $\mathbf{x}_1$ and condition $\mathbf{c}$};
        \node (noise) [block, below=of start] {Sample noise $\mathbf{x}_0$};
        \node (interpolate) [block, below=of noise] {Interpolate $\mathbf{x}_1$ to get $\mathbf{x}_{t'}$};
        
        % Loss computation
        \node (cond) [block, below=of interpolate, xshift=-2.0cm] {Compute conditional \\ loss $\mathcal{L}_{\text{cond}}$};
        \node (uncond) [block, below=of interpolate, xshift=2.0cm] {Compute unconditional \\ loss $\mathcal{L}_{\text{uncond}}$};
        
        \node (compare) [decision, below=of cond, xshift=2.0cm] {$\mathcal{L}_{\text{cond}} > \mathcal{L}_{\text{uncond}}$?};

        % Decision outcomes
        \node (uncond_branch) [block, below=of compare, xshift=-1.5cm, fill=red!15] {Unconditional \\ training with $\mathbf{x}_1$};
        \node (cond_branch) [block, below=of compare, xshift=1.5cm] {Ordinary training \\ with $\mathbf{x}_1$ and $\mathbf{c}$};

        \node (end) [block, below=of uncond_branch, xshift=1.5cm, yshift=-0.1cm] {Optimize model parameters $\theta$};

        % Lines
        \draw [line] (start) -- (noise);
        \draw [line] (noise) -- (interpolate);
        
        \draw [line] (interpolate) -| (cond);
        \draw [line] (interpolate) -| (uncond);
        
        \draw [line] (cond) -- (compare);
        \draw [line] (uncond) -- (compare);
        
        \draw [line] (compare) -- node[left, pos=0.3, xshift=-0.2cm] {Yes} (uncond_branch);
        \draw [line] (compare) -- node[right, pos=0.3, xshift=0.2cm] {No} (cond_branch);

        \draw [line] (uncond_branch) -- (end);
        \draw [line] (cond_branch) -- (end);

        \end{tikzpicture}
    \end{adjustbox}
    \caption{Flow diagram illustrating the training process with SPFM.} 
    % The process evaluates the efficacy of a given condition $\mathbf{c}$ by comparing the conditional loss ($\mathcal{L}_{\text{cond}}$) with the unconditional loss ($\mathcal{L}_{\text{uncond}}$). If the condition is determined to be ineffective ($\mathcal{L}_{\text{cond}} > \mathcal{L}_{\text{uncond}}$), the system reverts to unconditional learning.}
    \label{fig:flowchart}
\end{figure}

% We propose to exploit the conditional and unconditional vector fields to identify unreliable labels on-the-fly during the training of flow-matching models. Our intuition is that the flow matching loss is generally lower when a proper condition $\mathbf{c}$ is given compared to the one computed without the condition. In other words, our proposed method, SPFM, compares conditional and unconditional losses, $\mathcal{L}_{\text{cond}}$ and $\mathcal{L}_{\text{uncond}}$, respectively, and identify the condition \mathbf{c} as unreliable when $L_{cond} >L_{uncond.}$.

We propose SPFM, a method that leverages the conditional and unconditional vector fields to detect unreliable labels during the training of flow-matching models. Our key intuition is that the flow-matching loss should generally be lower when a correct condition $\mathbf{c}$ is provided than when it is omitted. Following this principle, SPFM compares conditional and unconditional losses, $\mathcal{L}_{\text{cond}}$ and $\mathcal{L}_{\text{uncond}}$, to identify potentially mislabeled or suspicious samples.

Formally, for a given data pair $(\mathbf{x}_1$, $\mathbf{c})$, SPFM computes the following losses:
\begin{align}
\mathcal{L}_{\text{cond}} = \big\| \mathbf{v}_\theta(\mathbf{x}_{t'}, t', \mathbf{c}) - (\mathbf{x}_1 - \mathbf{x}_0) \big\|^2, \\
\mathcal{L}_{\text{uncond}} = \big\| \mathbf{v}_\theta(\mathbf{x}_{t'}, t', \varnothing) - (\mathbf{x}_1 - \mathbf{x}_0) \big\|^2,
\end{align}
where $t' \in [0,1]$ denotes the interpolation time, and $\mathbf{x}_{t'}$ is the corresponding interpolated point between $\mathbf{x}_{0}$ and $\mathbf{x}_1$. The same $t'$ and $\mathbf{x}_0$ are used in both computations. If $\mathcal{L}_{\text{cond}} > \mathcal{L}_{\text{uncond}}$, SPFM flags the condition $\mathbf{c}$ as potentially unreliable and trains the model on $\mathbf{x}_1$ in an unconditional manner. Otherwise, training proceeds as in the standard flow-matching pipeline. In practice, SPFM is applied only after a warm-up period to mitigate false detections arising from uninformative early losses. Also, the interpolation time $t'$ is typically set to 0.5. The overall procedure is summarized in Fig.~\ref{fig:flowchart}.

SPFM offers significant advantages for learning with noisy datasets. For instance, it identifies potentially incorrect labels without relying on pretrained models, complex preprocessing, or extra modules. This makes the approach simple and easy to integrate. Moreover, SPFM can be seamlessly incorporated into existing flow-matching-based conditional generative models, providing a principled way to filter out low-quality training data. By self-purifying the training process, SPFM ensures that the model learns primarily from reliable samples, resulting in more robust and accurate generative performance. These properties set SPFM apart from methods that require external tools or auxiliary objectives.

\section{Synthetic Experiments}
\begin{figure}[t]
  \centering
  \begin{subfigure}{\columnwidth}
    \centering
    \includegraphics[width=0.95\linewidth]{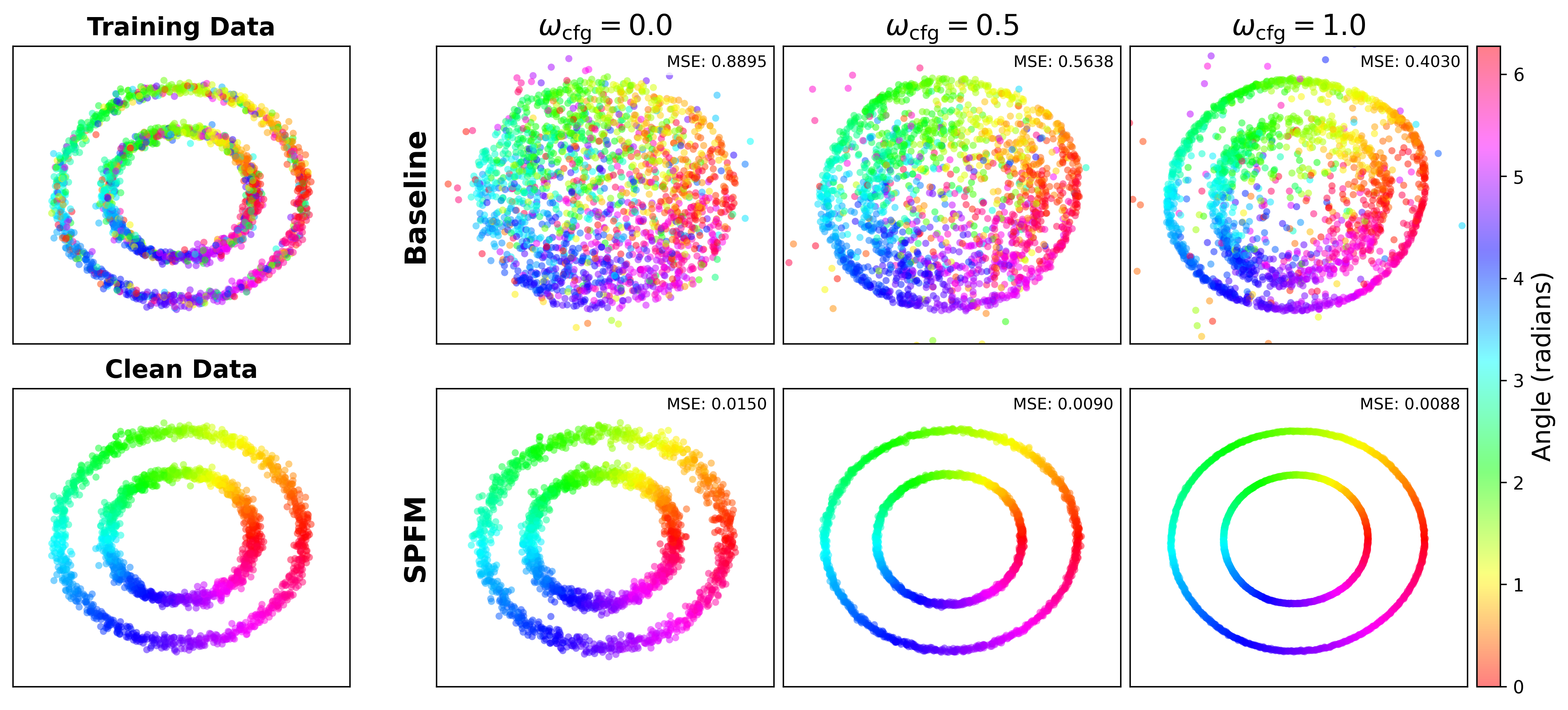}
  \end{subfigure}
  \begin{subfigure}{\columnwidth}
    \centering
\includegraphics[width=0.95\linewidth]{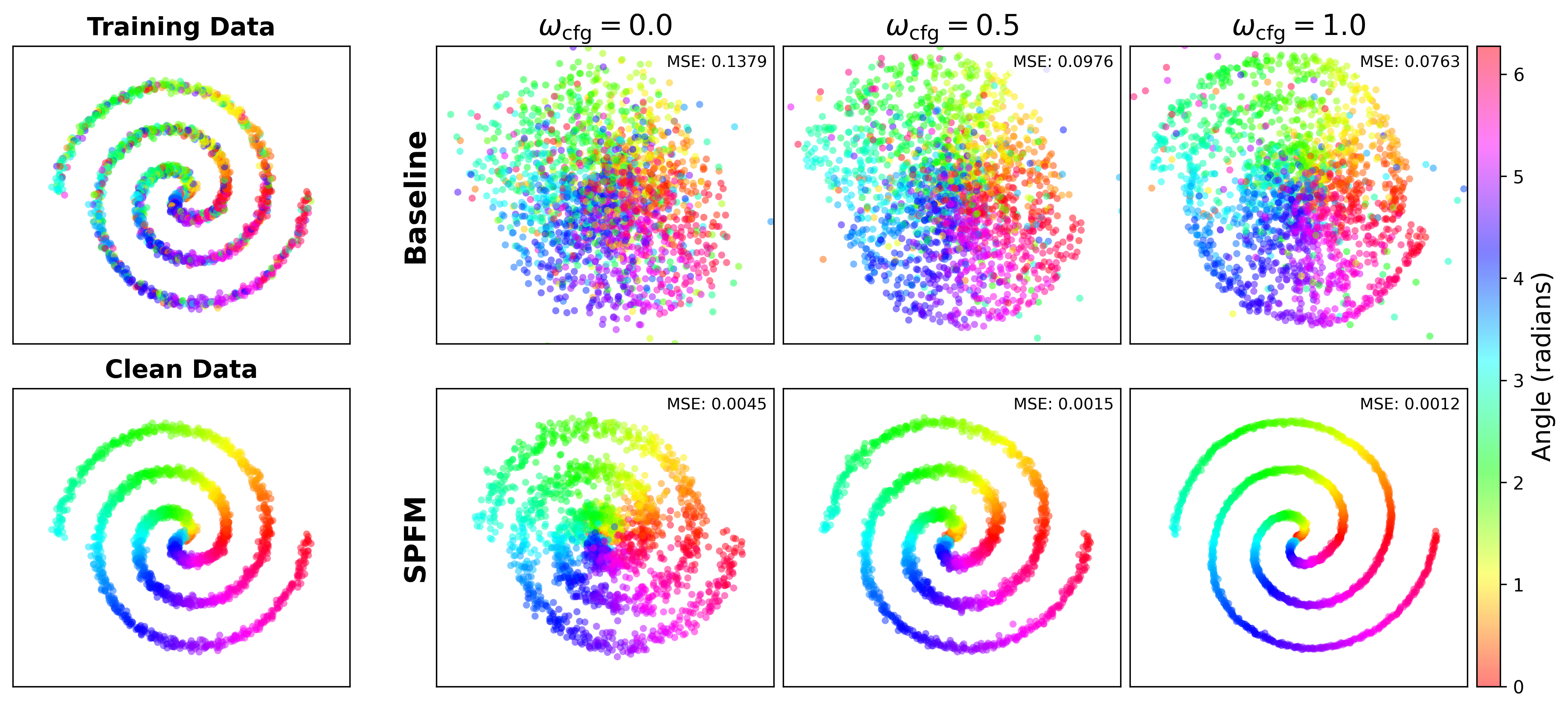}
  \end{subfigure}
  \caption{Comparative results on 2D synthetic datasets. Training data contains 40\% polar label noise, while clean data is noise-free. Colors indicate angular coordinates used for conditioning.}
  \label{fig:artificial_data}
\end{figure}

\subsection{Experimental setup}
We evaluate SPFM on two-dimensional artificial datasets to demonstrate its effectiveness in handling noisy labels. Specifically, we consider two synthetic datasets: two-circle and spiral shapes, where each sample is annotated with corresponding polar coordinates (angle and radius). To simulate noisy labeling scenarios, we corrupt 40\% of the labels by randomly reassigning polar coordinate conditions. This controlled setup enables a precise assessment of the impact of label noise on conditional generation quality.

We trained two flow-matching models with simple linear architectures: one equipped with SPFM and one without. During training, a dropout rate of 10\% was applied to the polar condition to enable classifier-free guidance. The self-purification mechanism was activated after a warm-up of 4 epochs within the total 100 epochs of training. For reproducibility, we release the code used in our experiments.\footnote{https://github.com/supertone-inc/self-purifying-flow-matching}

\subsection{Results}
Fig.~\ref{fig:artificial_data} presents the experimental results across different guidance scales. To quantitatively assess the sample quality, we measured the mean squared error (MSE) between the conditioned Euclidean point and the generated sample. It can be observed that SPFM consistently outperforms the baseline in all configurations, generating high-fidelity samples. Without guidance ($\omega_{\text{cfg}}=0.0$), the baseline model fails to capture the target distribution shapes, producing scattered and noisy samples that deviate from the given conditions. In contrast, SPFM generates well-structured samples that adhere to the angular condition, even recovering clean distributions that were not presented during training. As the guidance scale increases from 0.0 to 1.0, the baseline’s sample quality gradually improves, but it still underperforms and frequently violates the condition. On the other hand, SPFM consistently produces sharp, clean shapes that respect the conditioning, demonstrating its ability to capture the intrinsic relationship between data and polar coordinates even under noisy supervision. These results confirm that SPFM is an effective approach for training conditional flow-matching models in the presence of noisy labels.

\section{Text-to-Speech Experiments}
\begin{table}[t]
\caption{Performance comparison on TITW-KSKT for TTS models trained with TITW-Easy.}
\centering
\begin{tabular}{lccc}
\toprule
Model & UTMOS        & DNSMOS       & WER(\%)      \\ \midrule
TransformerTTS       & 2.06          & 2.50          & 24.90         \\
MQTTS                & 3.08          & 2.83          & 23.30         \\
GradTTS              & 2.18          & 2.39          & 11.90         \\
VITS                 & 2.77          & 2.74          & 53.00         \\ \midrule
SupertonicTTS        & \textbf{3.43 \scriptsize{± 0.01}} & 2.84 \scriptsize{± 0.01}          & 6.68          \\ 
+ SPFM & \textbf{3.43 \scriptsize{± 0.01}} & \textbf{2.86 \scriptsize{± 0.01}} & \textbf{5.96} \\ \bottomrule
\end{tabular}
\label{table:titw-easy}
\end{table}

\begin{table}[t]
\caption{Performance comparison on TITW-KSKT for TTS models trained with TITW-Hard.}
\centering
\begin{tabular}{lccc}
\toprule
Model & UTMOS        & DNSMOS       & WER(\%)      \\ \midrule
GradTTS              & 1.29          & 1.47          & 26.20         \\
VITS                 & 2.48          & 2.69          & 59.50         \\ \midrule
SupertonicTTS        & 3.50 \scriptsize{± 0.01}          & 2.88 \scriptsize{± 0.01}          & 7.60          \\ 
+ SPFM & \textbf{3.55 \scriptsize{± 0.01}} & \textbf{2.91 \scriptsize{± 0.01}} & \textbf{6.86} \\ \bottomrule
\end{tabular}
\label{table:titw-hard}
\end{table}

To validate the effectiveness of SPFM in real-world scenarios, we evaluate text-to-speech (TTS) models trained on the TITW database~\cite{jung25c_interspeech}. The TITW database contains two splits: TITW-Easy and TITW-Hard. TITW-Hard is an automatically annotated and segmented dataset derived from VoxCeleb1~\cite{Nagrani17} using WhisperX~\cite{bain23_interspeech}. TITW-Easy is a refined subset of TITW-Hard, where each utterance is enhanced using DEMUCS~\cite{defossez20_interspeech} and samples with a DNSMOS rating~\cite{reddy2021dnsmos} below 3.0 are excluded. Because the TITW database is collected in the wild, it is inherently noisy, making it a challenging and suitable benchmark for training robust TTS models under noisy labels.

\subsection{Experimental setup}
\begin{figure*}[t]
  \centering
  \includegraphics[width=\textwidth]{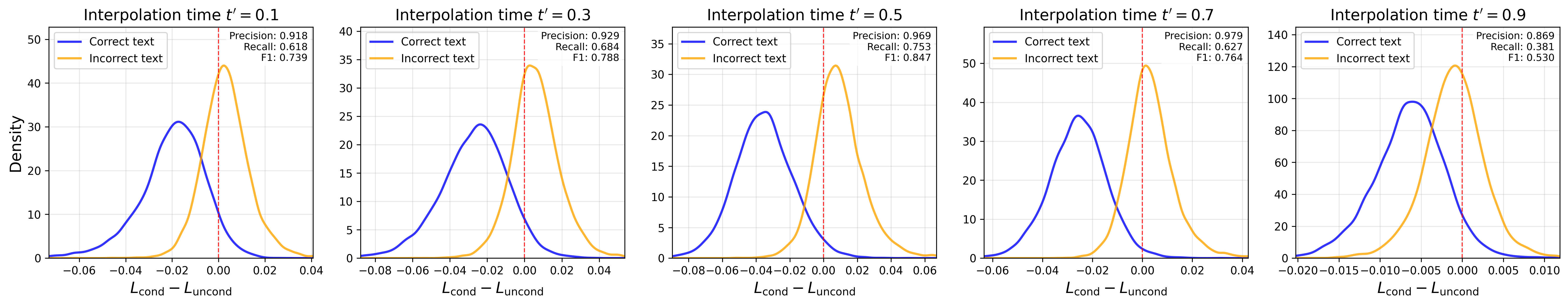}
  \caption{Distribution of difference between $\mathcal{L}_{\text{cond}}$ and $\mathcal{L}_{\text{uncond}}$ across different interpolation times.}
  \label{fig:wrong_cond}
\end{figure*}

We adopt SupertonicTTS~\cite{kim2025supertonictts, kim2025lengthawarerotarypositionembedding} as our baseline system due to its state-of-the-art performance and lowest pronunciation error rate on standard zero-shot TTS benchmarks. Specifically, we trained its text-to-latent modules both with and without SPFM for text conditioning. All models were optimized with the AdamW optimizer \cite{loshchilov2018decoupled}, using a batch size of 128 on four NVIDIA RTX 4090 GPUs and a batch expansion factor of 8. The initial learning rate was set to \num{5e-4} and was halved every 300k iterations over a total of 700k training iterations. SPFM was applied after a warm-up period of 40k iterations.

Evaluation was conducted on TITW-KSKT~\cite{jung25c_interspeech}. Since this test set was unavailable at the time of writing, we constructed evaluation samples following the procedure in \cite{jung25c_interspeech}. Specifically, we collected 9,113 utterances from 40 speakers in accordance with the VoxCeleb1-O evaluation protocol. For each sample, a separate utterance from the same speaker was randomly selected as reference speech. 

We report three evaluation metrics: UTMOS~\cite{saeki22c_interspeech}, DNSMOS~\cite{reddy2021dnsmos}, and word error rate (WER). UTMOS and DNSMOS serve as proxies for perceptual speech quality, while WER quantifies pronunciation accuracy and intelligibility by comparing synthesized speech to the target text. WER was computed by transcribing the synthesized speech with Whisper large-v2~\cite{radford2023robust} and comparing against the ground-truth text. All metrics were calculated using the VERSA toolkit~\cite{shi2024espnet}.

\subsection{Performance evaluation}
Table \ref{table:titw-easy} summarizes the results on the TITW-KSKT benchmark for models trained with TITW-Easy. SupertonicTTS already demonstrates strong performance, outperforming other baseline models such as TransformerTTS~\cite{li2019neural}, MQTTS~\cite{chen2023vector}, GradTTS~\cite{popov2021grad}, and VITS~\cite{kim2021conditional}. Incorporating SPFM further improves performance, notably reducing WER from 6.68\% to 5.96\% and slightly increasing DNSMOS from 2.84 to 2.86, while maintaining the highest UTMOS of 3.43.

On the more challenging TITW-Hard split, which contains noisier recordings, the benefits of SPFM are even more pronounced. As shown in Table~\ref{table:titw-hard}, SupertonicTTS achieves a UTMOS of 3.50, DNSMOS of 2.88, and WER of 7.60\%. With SPFM, all metrics improve to a UTMOS of 3.55, DNSMOS of 2.91, and WER of 6.86\%. Overall, these results confirm that SPFM consistently enhances TTS performance, particularly when training data are noisy, establishing a new benchmark on the TITW dataset.

\subsection{Analysis of retained and filtered data under SPFM}

\begin{table}[t]
\caption{Evaluation of dataset quality before and after self-purification. `Retained' indicates the subset of samples remaining after purification, while `Filtered' refers to samples identified as unreliable.}
\centering
\begin{tabular}{clccc}
\toprule
TITW                    & Data     & \#Sample  & UTMOS & WER(\%) \\ \midrule
\multirow{3}{*}{Easy} & Original & 10,000 & 3.41 \scriptsize{± 0.01}  & 11.53   \\ \cmidrule(l){2-5} 
                           & Retained & 9,700 & 3.42 \scriptsize{± 0.01}  & 11.24   \\
                           & Filtered & 300 & 3.19 \scriptsize{± 0.07}  & 24.58   \\ \midrule
\multirow{3}{*}{Hard} & Original & 10,000 & 3.06 \scriptsize{± 0.01}  & 11.77   \\ \cmidrule(l){2-5} 
                           & Retained &9,763 & 3.07 \scriptsize{± 0.01}  & 11.52   \\
                           & Filtered  & 237 & 2.75 \scriptsize{± 0.10}  & 26.38   \\ \bottomrule
\end{tabular}
\label{table:dataset-quality}
\end{table}

% \begin{figure}[t]
%     \centering
%     \includegraphics[width=1.0\columnwidth]{figures/utmos_combined_density.png}
%     \caption{Distribution of UTMOS scores for remaining and filtered samples by self-purification.}
%     \label{fig:utmos}
% \end{figure}

To better understand the effect of SPFM, we analyzed the characteristics of the retained and discarded samples under the proposed mechanism. From each of TITW-Easy and TITW-Hard, we randomly selected 10k samples and classified them using the pretrained model. 

Table \ref{table:dataset-quality} presents the results of data quality assessment. Notably, the filtered subsets exhibit substantially lower quality than the retained ones. For instance, in the TITW-Easy split, filtered samples achieve a UTMOS of 3.19 and a WER of 24.58\%, whereas the retained subset reaches with a UTMOS of 3.42 and a WER of 11.24\%. A similar trend is observed for TITW-Hard, where the filtered subset records a UTMOS of 2.75 and a WER of 26.38\% , whereas the retained subset achieves 3.07 and 11.52\%. These results indicate that SPFM effectively identifies and removes samples with low perceptual quality or high transcription noise, thereby improving the reliability of the training set while retaining the majority of the data.

% Figure \ref{fig:utmos} further illustrates these findings by showing the UTMOS distribution of the two subsets. The filtered samples are consistently left-shifted relative to the remaining data across both TITW-Easy and TITW-Hard. This indicates that the filtered subset contains many utterances with poor perceptual fidelity. Since WER is measured by comparing transcriptions from Whisper large-v3-turbo against the ground-truth text, the significantly degraded WER of filtered samples suggests either (i) mislabeled or noisy annotations, or (ii) speech segments that are too degraded or challenging for reliable learning.

% These analyses confirm that SPFM effectively identifies and removes data of low perceptual quality or high transcriptional noise, thereby improving the reliability of the training corpus without discarding a large fraction of samples.

\subsection{Effect of label correctness on loss difference}
We further analyzed the distribution of the difference between the conditional and unconditional losses, \(\mathcal{L}_{\text{cond}} - \mathcal{L}_{\text{uncond}}\), computed at various interpolation times 
\(t' = 0.1, 0.3, 0.5, 0.7, 0.9\). Both losses were calculated using the pretrained model on the TITW-Easy set. To simulate noisy labels, the model was provided with incorrect text that did not match the given speech, whereas for correct label simulation, the ground-truth transcription was used.

As shown in Figure \ref{fig:wrong_cond}, the distributions of loss difference consistently shift to the right across all interpolation times when incorrect text is used. This supports our hypothesis that \(\mathcal{L}_{\text{cond}} - \mathcal{L}_{\text{uncond}}\) is a reliable signal for label mismatches. Notably, the distributions are most discriminative at an interpolation time of $t'=0.5$, achieving the highest F1-score of 0.847. At this setting, SPFM is particularly effective at identifying mislabeled samples while minimizing false detections. For lower $t'$, the interpolated data is overly noisy, whereas at $t'$ values approaching 1, the model can infer accurately without strong reliance on the condition, diminishing the impact of an incorrect label.

\section{Limitations}
This study identifies unreliable labels using the criterion $L_{\text{cond}} - L_{\text{uncond}} > 0$ at a fixed timestep $t'$. Alternative approaches, such as adaptive thresholds or evaluation across multiple timesteps, could potentially improve detection performance, but exploring these methods is beyond the scope of this paper. In addition, our experiments primarily focus on TTS. Since SPFM is generally applicable to conditional flow-matching models, evaluating its performance on a broader range of real-world datasets is necessary to fully understand its generalization capabilities and scalability across different domains.

\section{Conclusion}
In this work, we introduced a novel and principled method, namely SPFM, for filtering unreliable data during the training of conditional flow-matching models. Our approach leverages the model's own functionality, identifying suspicious labels on-the-fly by comparing conditional and unconditional losses. A key advantage of SPFM is its simplicity and ease of integration, as it requires no auxiliary modules, pretrained models, or complex preprocessing pipelines. Our experiments on both synthetic and real-world datasets demonstrated that SPFM effectively mitigates the impact of label noise, leading to improved performance and establishing a new state-of-the-art benchmark on the TITW dataset. We believe our findings will enable more trustworthy and high-performing generative models in real-world applications.

% Our experiments on synthetic 2D datasets demonstrated that SPFM effectively mitigates the challenge of label noise, enabling the model to generate samples that closely adhere to the specified conditioning, whereas standard models fail. Furthermore, we validated our method in a real-world scenario by applying it to TTS models with the TITW database. The results showed that SPFM consistently improves performance across all metrics, particularly with reduced WER, and establishes a new state-of-the-art benchmark on this dataset. Our analysis further confirmed that SPFM successfully identifies and filters samples with low perceptual quality and high transcription noise, thereby enhancing the overall reliability of the training data. SPFM offers a practical and effective solution for robustly training conditional generative models on large-scale, imperfect datasets.

\pagebreak
% \clearpage

\bibliographystyle{IEEEbib}
\bibliography{refs}

\end{document}